\documentclass[twocolumn,tighten]{aastex6}
\usepackage{hyperref,url}
\usepackage{natbib,graphicx,amsmath}

\shorttitle{Delta Scuti in NGC 1846}
\shortauthors{Salinas et al.}

\newcommand{\dsct}{$\delta$~Sct\,}
\begin{document}

\title{Stellar variability at the main-sequence turnoff of the intermediate-age LMC cluster NGC 1846$^{\dagger}$}

\author{R. Salinas\altaffilmark{1}, M. A. Pajkos\altaffilmark{2,7}, A. K. Vivas\altaffilmark{3}, J. Strader\altaffilmark{4} and R. Contreras Ramos\altaffilmark{5,6} }
          
\altaffiltext{1}{Gemini Observatory, Casilla 603, La Serena, Chile; rsalinas@gemini.edu}
\altaffiltext{2}{Department of Physics and Astronomy, Butler University, Indianapolis, IN 46208, USA}
\altaffiltext{3}{Cerro Tololo Interamerican Observatory, National Optical Astronomy Observatory, Casilla 603, La Serena, Chile}
\altaffiltext{4}{Department of Physics and Astronomy, Michigan State University, East Lansing, MI 48824, USA}    
\altaffiltext{5}{Millennium Institute of Astrophysics, Av. Vicu\~na Mackenna 4860, 782-0436 Macul, Santiago, Chile}
\altaffiltext{6}{Instituto de Astrof{\'i}sica, Pontificia Universidad Cat\'olica de Chile, Av. Vicu\~na Mackenna 4860, 782-0436 Macul, Chile}

\begin{abstract}

Intermediate-age star clusters in the LMC present extended main sequence turnoffs (MSTO) that have been attributed to either multiple stellar populations or an effect of stellar rotation. Recently it has been proposed that these extended main sequences can also be produced by ill-characterized stellar variability. Here we present Gemini-S/GMOS time series observations of the intermediate-age cluster NGC 1846. Using differential image analysis, we identified 73 new variable stars, with 55 of those being of the Delta Scuti type, that is, pulsating variables close the MSTO for the cluster age. Considering completeness and background contamination effects we estimate the number of \dsct belonging to the cluster between 40 and 60 members, although this number is based on the detection of a single \dsct within the cluster half-light radius. This amount  of variable stars at the MSTO level will not produce significant broadening of the MSTO, albeit higher resolution imaging will be needed to rule out variable stars as a major contributor to the extended MSTO phenomenon. Though modest, this amount of \dsct makes NGC 1846 the star cluster with the highest number of these variables ever discovered. Lastly, our results are a cautionary tale about the adequacy of shallow variability surveys in the LMC (like OGLE) to derive properties of its \dsct population.
\end{abstract}

   
\section{Introduction}\label{sec:intro}

\let\thefootnote\relax \footnotetext{$^{\mathrm{\dagger}}$Based on observations obtained at the Gemini Observatory, which is operated by the Association of Universities for Research in Astronomy, Inc., under a cooperative agreement with the NSF on behalf of the Gemini partnership: the National Science Foundation (United States), the National Research Council (Canada), CONICYT (Chile), Ministerio de Ciencia, Tecnolog\'{i}a e Innovaci\'{o}n Productiva (Argentina), and Minist\'{e}rio da Ci\^{e}ncia, Tecnologia e Inova\c{c}\~{a}o (Brazil).}

\let\thefootnote\relax \footnotetext{$^{7}$CTIO/Gemini REU student.}

Intermediate-age (IA) star clusters in the Large Magellanic Cloud (LMC) exhibit extended main sequence turn-offs (MSTOs) inconsistent with single stellar populations \citep{mackey07,mackey08,milone09}, which cannot be explained by photometric errors, contamination from the LMC field or binaries \citep[e.g.][]{goudfrooij09}. These extended MSTOs can be interpreted as two bursts of star formation separated by a few hundred Myr or a continuous star formation lasting a similar amount of time \citep{mackey07}. This interpretation, however, is complicated because any age spread at MSTO level should also  be visible at the red clump, but the morphology of red clumps is rather consistent with single stellar populations \citep[][but see \citealt{goudfrooij15} for a different view]{li14,bastian15}. Moreover, for younger clusters with ages of a few Myrs, where any extended star formation should be even clearer in their CMDs, the evidence of departures from single stellar populations remains highly debated \citep{bastian13b,correnti15,niederhofer15,milone17}.

An alternative explanation could be given by stellar rotation. Fast-rotating stars with ages $\sim$ 1.5 Gyr, will have different temperatures as a function of latitude. When viewed from different angles, these temperature differences will be seen as a range of colors and luminosities,  producing an extension to the MSTO, mimicking the effect of multiple stellar populations \citep{bastian09,yang13,brandt15}, although it has been claimed that rotation alone cannot fully reproduce the extended MSTO morphology \citep{girardi11,goudfrooij17}.

Recently, \citet{salinas16b} have shown that another previously overlooked factor must be considered to understand these extended MSTOs. The instability strip will cross the upper MS and MSTO area for clusters with ages between 1 to 3 Gyr and therefore a certain number of the stars within the instability strip will develop pulsations. These main-sequence pulsators are known as Delta Scuti stars \citep[hereafter \dsct, see e.g.][for a review]{breger00}. For CMDs obtained using single images per filter, as the great majority of CMDs derived from \textit{HST} images \citep{mackey07,mackey08,milone09}, as well as most of the ground-based observations \citep[e.g.][]{piatti14}, the act of observing these variables at a random phase means their magnitudes and colors will be away from their static values, producing an artificial broadening of the MSTO that can be misinterpreted as the effect of an extended star formation history or rotation.

The impact of variables near the MSTO will depend on the percentage of stars developing pulsations (the incidence) and on the magnitude of the pulsation amplitudes. These factors are very poorly constrained in extragalactic systems. In Carina, for example, \citet{vivas13} find a lower limit of 8\% for  the incidence and an amplitude distribution with a peak at $A_V\sim0.5$ mag.

The properties of these quantities in the LMC clusters are even less constrained. That is the case because it is difficult to detect them. First, they are faint stars. With magnitudes between 20 and 22 at the LMC distance, they are out of reach of most of the large variability surveys which are conducted with small-aperture telescopes. Second, their periods are short which has a consequence that the exposure times must kept short to sample correctly the light curve. At least medium size telescopes are needed then.

A quick revision through the catalog of \dsct in the LMC of \citet{poleski10} reveals that for the 14 IA clusters listed in \citet{piatti14}, between zero and two \dsct are found per cluster, indicating that crowding significantly hampers the reliability of OGLE at these faint magnitudes, and only a handful of \dsct have been found in other searches \citep[e.g.][]{kaluzny03}.
 
\subsection{The intermediate-age cluster NGC 1846}

NGC 1846 (RA=05:07:34.9, Dec=-67:27:32.45) is a rather massive \citep[$M=1.25\times10^5$M$_\odot$;][]{baumgardt13}, intermediate-age \citep[$\sim$2 Gyr;][]{mackey07} and metal-rich \citep[$\lbrack$Fe/H$\rbrack$=--0.49;][]{grocholski06} LMC cluster. It was the first LMC cluster where an extended and bifurcated  MSTO was detected and firmly established \citep{mackey07,mackey08,goudfrooij09}, and where explanations involving field contamination and binary evolution were discarded as a cause \citep{goudfrooij09}.

With a core radius of $\sim$ 6.5 pc \citep{keller11}, NGC 1846 is also one of the most extended IA clusters, which makes it a more suitable candidate for ground-based photometry. Large core radii have been suggested to be associated with the presence of extended MSTOs \citep{keller11}.

We assess the role of \dsct in the morphology of the MSTO in IA clusters in the LMC using new time series imaging of the LMC cluster NGC 1846.

\begin{figure}
   \centering
   \includegraphics[width=0.47\textwidth]{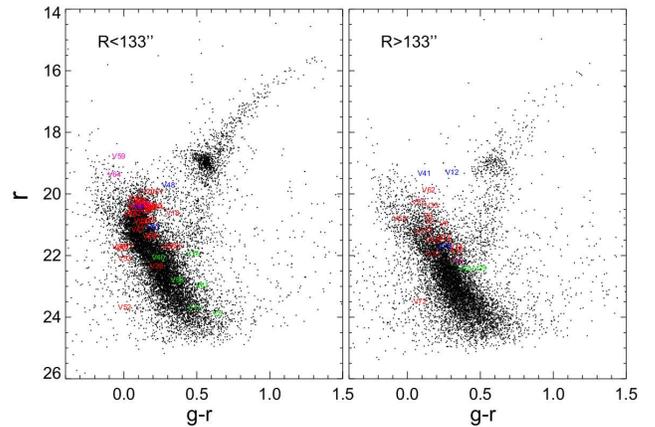}
      \caption{GMOS photometry of the NGC 1846 field. Left panel shows the inner 133\arcsec\, from the cluster center, while the right panel shows the area outside this limit out to the edges of the GMOS fov. The 133\arcsec\, limit defines equal areas within the fov. Variable stars discovered in the field are colored according to their classification: delta Scuti (red), RR Lyrae (blue), eclipsing (green) and unknown classification (magenta). The inner population of \dsct more than doubles the one on the outer field.}
         \label{fig:cmd}
   \end{figure}

\section{Observations and data reduction}\label{sec:data}
   
Observations of NGC~1846 were conducted using the 8.1m Gemini South telescope, located at Cerro Pach\'on, Chile, on the night of December 30, 2015 under the Gemini Fast Turnaround mode (Gemini program GS-2015B-FT-7).  The imaging mode of the Gemini Multi-Object Spectrometer \citep[GMOS,][]{hook04} provided us a 5.5 square arcminute field of view (fov). The SDSS filter system was used to yield 6, 66 and 6 images, in $g$, $r$, and $i$, each having exposure times of 120 s, 120 s, and 90 s, respectively.  The total time span of observations was 0.136 days (3.26 hours); adequate for detection of \dsct which will have periods of less than $\sim6$ hours. The GMOS-S array detector consists of three 2048$\times$4176 pixels Hamamatsu detectors, each separated by a gap $\sim$\~30 pixels wide. Observations were obtained with 2$\times$2 binning, resulting in a pixel scale of 0.16\arcsec pixel$^{-1}$.

  \begin{figure*}[t]
   \centering
   \includegraphics[width=\textwidth]{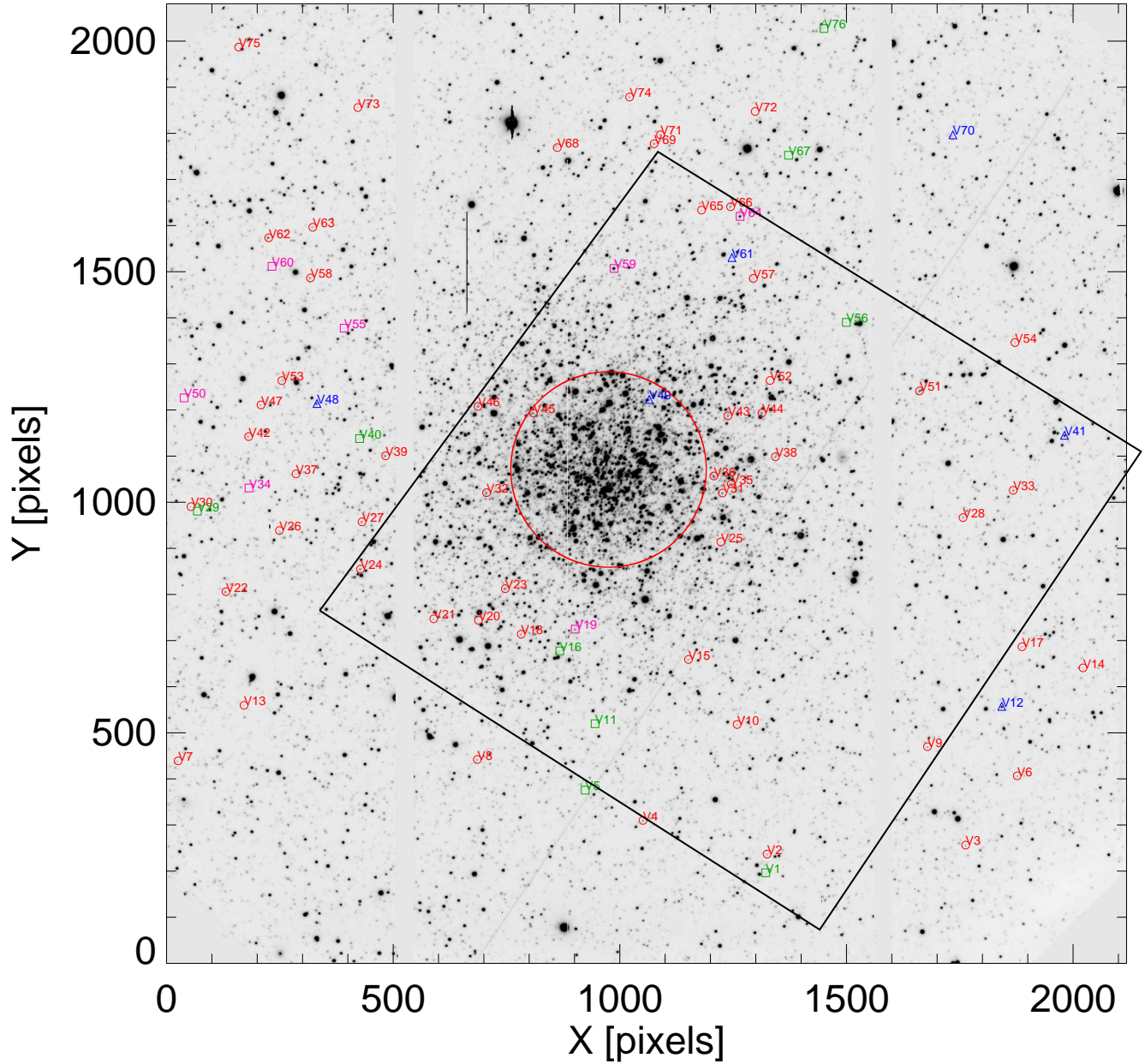}
      \caption{A finder chart for variables in the NGC 1846 field based on a GMOS $r$ image.  North is up and East to the left. Positions of \dsct are indicated with red circles, RRL, with blue triangles, eclipsing binaries with green squares. Magenta squares mark the position of variables of unknown type. The central big red circle represents the half-light radius from \citet{goudfrooij09}. The fov of GMOS is approximately $5.5\times5.5$ arcmin, while the big black square indicates the \textit{HST}/ACS pointing used by \citet{goudfrooij09} and \citet{milone09}. North is up, East to the left.}
         \label{fig:finder}
   \end{figure*}
   
Raw data retrieved from the Gemini Observatory  Archive$^1$\footnote{$^1$\url{https://archive.gemini.edu}}, were reduced using the \textsc{gemini} package in \textsc{iraf} $^2$\footnote{$^2$ IRAF is distributed by the National Optical Astronomy Observatories, which are operated by the Association of Universities for Research in Astronomy, Inc., under cooperative agreement with the National  Science Foundation.}.  Specifically, the \textsc{gmos} subpackage allowed us to bias and flatfield correct the raw images, as well as mosaicing the chips and trimming their overscan region. Image quality was measured with the \textsc{gemseeing} task. The median FWHM for the $r$ dataset was 0.8\arcsec.
   
   \subsection{Photometry}

Photometry of the images was obtained using the \textsc{daophot/allstar/allframe} suite of programs developed by \citet{stetson87,stetson94}. As a first step, \textsc{daophot} was run over all images. The preliminary positions and aperture-photometry magnitudes were used to obtain the coordinate transformations between the frames with the help of \textsc{daomatch/daomaster} \citep{stetson93}. Close to 50 bright isolated stars were visually chosen on the best seeing image of each filter to model the psf as a linearly varying Gaussian profile. The same psf stars were used in the rest of the frames transforming the coordinates using the \textsc{daomaster} output. Once psf-photometry was obtained for all images with \textsc{allstar}, a deep reference frame was constructed using the 20 best seeing images. This reference frame was used to obtain the positions of the stars to be measured by \textsc{allframe} \citep{stetson94}, which fits simultaneously the PSF to all stars in all the available images. Final catalogues with mean instrumental magnitudes in $g$, $r$ and $i$ measured by \textsc{allframe} were obtained with \textsc{daomaster}.

Given the absence of standards stars taken on the night of the observations, calibration to the standard system was achieved using the transformation equations provided by the observatory for the Hamamatsu CCDs$^3$. \footnote{$^3$\url{https://www.gemini.edu/?q=node/10445}}

The color-magnitude diagram of the NGC 1846 field can be seen in Fig. \ref{fig:cmd}. This diagram has further cleaning steps leaving only stars with \textsc{allstar} parameters \verb+chi+$<$ 10 and $-2<$\verb+sharp+ $< $2 and photometric errors less than 0.1 mag in each filter. Photometry of the field is shown split in two equal areas. The left panel shows the inner area ($R<133$\arcsec), where the MSTO of the cluster can be seen at $r\sim21$ and $g-r\sim 0.15$, and also evidence for old, intermediate and young populations on the LMC field. The subgiant branch of NGC 1846 is barely distinguishable at $r\sim20$ and $0.2<g-r<0.5$. The right panel shows the surrounding field of $R>133\arcsec$. Here the intermediate-age MSTO and red clump are less pronounced and even though a young and intermediate-age population still exist, the dominant feature is the old LMC population, whose MSTO is around $r=23$.  The image quality of ground-based observations does not permit to see the double MSTO, clearly seen with \textit{HST} observations \citep[e.g.][]{mackey07,milone09}.

\section{Variable stars in NGC 1846}\label{sec:variables}

\subsection{Known variables}

We searched for the known variables in the cluster using the on-line search tool provided by OGLE$^4$\footnote{$^4$\url{http://ogledb.astrouw.edu.pl/~ogle/CVS/}}. OGLE lists 44 variables within the field of view  observed with GMOS. Of these, 39 are classified as long-period variables, four as RR Lyrae (RRL) and one as Cepheid \citep{soszynski09a,soszynski09b}. No short-period variables were found in this cluster by OGLE. Given the age of the cluster, any RRL or Cepheid will not be a member of the cluster, but members of the LMC field.

\subsection{Searching for new variables}

New variables stars in the NGC 1846 field  were searched using the image subtraction package ISIS \citep[v 2.1,][]{alard00}. ISIS first registers images to a common astrometric system. Then a reference frame is constructed as a median from the best seeing images. This reference is then convolved to match the psf of the rest of the images. Once the psf is matched, the subtraction is applied. This approach leaves in principle only the variable sources as residuals in these subtracted images. ISIS also constructs a variance image as the mean of absolute normalized deviations. This variance image is then visually inspected for meaningful variations in order to discard spurious artifacts produced, for example, around saturated stars \citep[e.g.][]{salinas16}. ISIS finally makes psf photometry on the selected positions were variability is suspected, giving as output light curves in fluxes relative to the reference frame. Variable sources were searched in the $r$ dataset which had the higher cadence of observations. Relative flux light curves are transformed into magnitudes following the procedure from \citet{catelan13}.  Periodicity in the light curves was searched using the phase dispersion minimization algorithm \citep{stellingwerf78} as implemented in \textsc{iraf}, using as limiting periods 0.001 and 0.3 days 

   \begin{figure}
   \centering
   \includegraphics[width=0.47\textwidth]{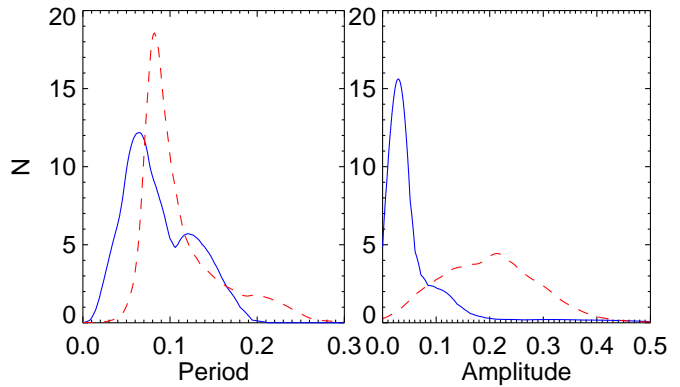}
      \caption{Period (left panel) and amplitude (right panlel) distribution of \dsct in NGC 1846 (blue solid lines) compared to \dsct in the LMC field from \citet{poleski10} (red dashed lines). Note that \citet{poleski10} amplitudes are in $I$ while our amplitudes are in $r$. The bump in the period distribution is an artifact produced by time span limit of the observations.   }
         \label{fig:pa}
   \end{figure}

\subsection{Variable star classification}\label{sec:classification}

The visual inspection of the variance image produced 164 candidate variable sources. 
Out of this initial list, after a careful visual inspection of the light curves, we settled on 76 genuine varying sources in the NGC 1846 field covered by GMOS. Two of these were not detected by DAOPHOT in either the $g$ nor the $r$ filters due to crowding, and therefore their light curves could not be transformed to magnitudes. Additionally, three more were only detected in the better quality $r$ data. Our final classification, based on the colors, magnitudes, periods, shapes of the light curve and amplitudes of each source, gives as result 55 \dsct, 8 eclipsing binaries, 5 RRL stars (3 of them already detected by OGLE) and 7 sources with no clear classification. 

Table \ref{table:variables} gives positions, periods and intensity-weighted magnitudes for all these variables. Given the short time span of the observations, for many variables only a lower limit for the period is provided.  Phased light curves can be seen in Figs. \ref{fig:variables1} to \ref{fig:variables4}. Additionally, Fig. \ref{fig:rrl} gives light curves for the candidate and known RRL stars in the field in Julian date versus magnitude. Appendix \ref{sec:notes} also gives notes for some of the variables, especially those with some ambiguity in their classification.

 \subsection{Period and amplitude distribution}
    
 Figure \ref{fig:pa} shows the period and amplitude distribution for 48 \dsct in NGC 1846 where both quantities have been measured, together with the same quantities for 937 \dsct in the LMC from \citet{poleski10}. Both distributions were obtained using an adaptive kernel density estimator \citep{epa69}.
 
The period distribution of \dsct in NGC 1846 (left panel) has a peak at $\sim$0.06d and a tail towards longer periods. The secondary peak around at $\sim$0.13d is an artifact arising from the time span of the observations, where this lower limit was assigned for \dsct with longer periods (see Table \ref{table:variables}). The peak of the period distribution from \cite{poleski10} is slightly higher, 0.08d, but significant.  The percentage of \dsct in NGC 1846 with $P<0.05$d is 20\% while in the \cite{poleski10} sample this is only 0.2\%. The scarcity of these short-period \dsct in the \cite{poleski10} sample may indicate that the cadence of OGLE observations, close to 3 days \citep{poleski10}, is inadequate to find them.
 
 The amplitude distribution of \dsct in NGC 1846 has its peak around 0.03 mag with a tail towards larger amplitudes up to 0.4 mag. Even though the OGLE data gives amplitudes in the $I$ filter instead of $r$, it is obvious that the amplitude distribution in the LMC field is very different to the one in NGC 1846. OGLE observations are most likely severely missing a large part of the \dsct in the LMC field. 
 
\section{The influence of Delta Scuti in the MSTO morphology}\label{sec:total}

The position of each variable in a CMD can be seen in Fig. \ref{fig:cmd}. As expected, \dsct cluster around the IA MSTOs of the cluster and the field population, with a few along the upper main sequence, and another group that are probably background \dsct. Field RRL appear as slightly brighter and redder than the IA TO. The faintness of RRL is partly because all their light curves miss the maximum, making their mean magnitudes dimmer.  

Fig. 2 in \citet{salinas16b} shows that hundreds or even thousands of \dsct would be needed to produce a significant broadening of the MSTO in IA clusters.

In order to compare with the predictions of \citet{salinas16b} we need to estimate the total number of \dsct in the cluster.

\subsection{Artificial stars and completeness}\label{sec:completeness}

\begin{figure}
   \centering
   \includegraphics[width=0.47\textwidth]{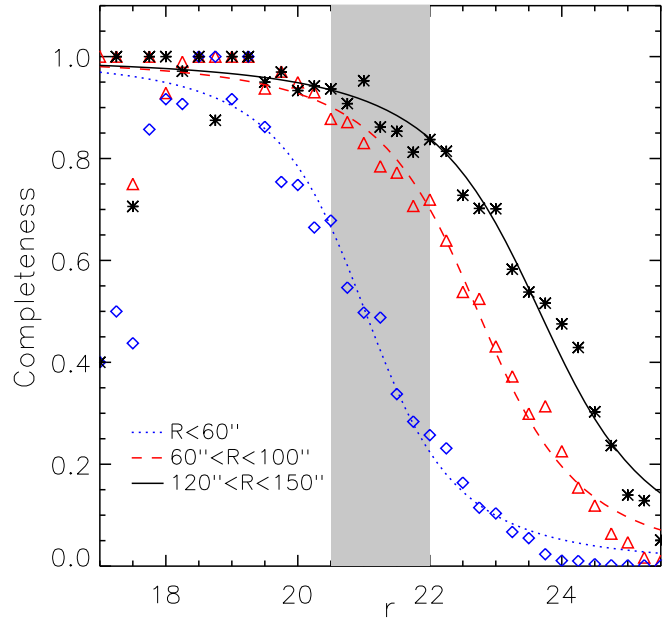}
      \caption{Results from the artificial stars experiment.  The different symbols show the completeness at different cluster-centric annuli. The grey stripe highlights the approximate magnitude range that \dsct will occupy at the cluster's distance.}
         \label{fig:complete}
   \end{figure}
   
   \begin{figure}
   \centering
   \includegraphics[width=0.47\textwidth]{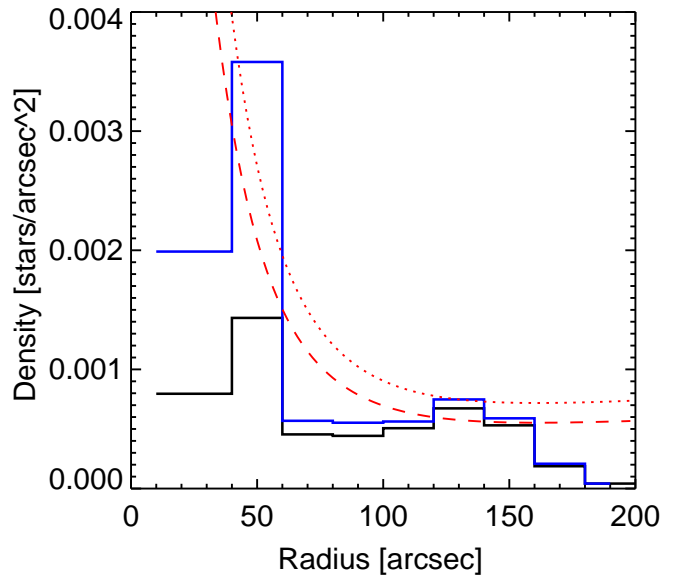}
      \caption{Radial distribution of the discovered \dsct. The black solid line shows the detected distribution, while the blue line shows the corrected distribution based on the completeness test. The dashed and dotted lines represent the scaled King profiles as described in the text. The estimated numbers for field and cluster \dsct come from the integration of these King profiles.}
         \label{fig:king}
   \end{figure}
   
To know how many detections we are missing given crowding and the image quality, we set up a standard artificial stars experiment using the task \verb+addstar+ within \textsc{daophot}. We added 45\,000 artificial stars (in 30 independent runs to avoid overcrowding) with magnitudes between $25<r<15$, whose $g-r$ colors and luminosity function were extracted from an IA synthetic CMD produced using BaSTI models \citep{pietrinferni04}. Stars were spatially distributed mimicking the distribution in the field, with 2/3 of the stars in each run randomly distributed and a third with a Gaussian distribution around the cluster center. Once the photometry over the frames with the added stars was done, artificial stars were considered as recovered if they lied within two pixels from their input positions and if the magnitude difference between output and input magnitudes was less than 0.5\,mag.

As seen in Fig, \ref{fig:complete}, completeness is greatly compromised in the inner $\sim 50$\arcsec, affecting at the 50\% level the 20$<r<$21.5 magnitude range where most of the \dsct lie. The completeness is close to 90\% for stars outside $\sim120$\arcsec\, in the same magnitude range. Lines represent the same interpolating function used e.g. in \citet{salinas15a}.

\subsection{Scaling to the total number}

Once we know how many \dsct we are missing as function of radius, we need to use this information to estimate the number of \dsct in the inner parts of the cluster where the completeness is poor. To this end we assume the incidence of \dsct will not vary with radius, and that \dsct follow the same radial distribution as the bright stars that dominate the overall light distribution, that is, there should be no significant mass segregation between the RGB stars and the upper MS stars, given their mass difference of less than $\sim 15$\%. The absence of a strong mass segregation between upper MS stars and RGB stars in NGC 1846 is confirmed by the analysis of \citet{goudfrooij09} based on \textit{HST} data.

\citet{goudfrooij09} fit the radial distribution of stars in NGC 1846 with a \citet{king62} profile

\begin{equation}
n(r) = n_0 \: \left( \frac{1}{\sqrt{1 + (r/r_c)^2}} - \frac{1}{\sqrt{1+c^2}}
 \right)^2 \; + \; {\rm bkg} 
\label{eq:king}
\end{equation}
finding best fit-parameters $n_0=8.1$, $r_c=26\arcsec$, $c=6.2$ and bkg=0.267.

Under the assumption that \dsct follow the distribution of light, we can scale this profile to find a total number of \dsct. We use the $120\arcsec<R<140$\arcsec\, range to define the normalization of the King profile for the \dsct. From the completeness experiment (Sec. \ref{sec:completeness}), a 90\% completeness for \dsct is expected at this distance. In this annulus there are 11 detected \dsct. We integrate the King profile between these limits and find a normalization factor of 0.0269 gives the expected number of 12.2 \dsct in the annulus. Integrating now the King profile from 0 to the tidal radius \citep[$\sim200\arcsec$,][]{goudfrooij09} we obtain 150 \dsct, where 90 would correspond to the background and 60 would be cluster members. 

This number is sensitive to the choice of annulus. If we select instead the $100\arcsec<R<120$ \arcsec\, range (where we are approximately 80\% complete), the same exercise gives us a total number of 45 \dsct members, which indicates that extrapolations, given our severe inner incompleteness, are necessarily very uncertain. Both King profiles can be seen in Fig. \ref{fig:king}.

Another uncertainty comes from the background level. Structural parameters of NGC 1846 measured by \citet{goudfrooij09} using ACS data might overestimate the background given the limited field of view of the ACS. If we assume 90\% of the  background given by \citet{goudfrooij09} then the total number of \dsct increases to a range between 50 and 65. At 70\% of the background, this range increases to 62 and 84.

\begin{figure}[t]
   \centering
   \includegraphics[width=0.48\textwidth]{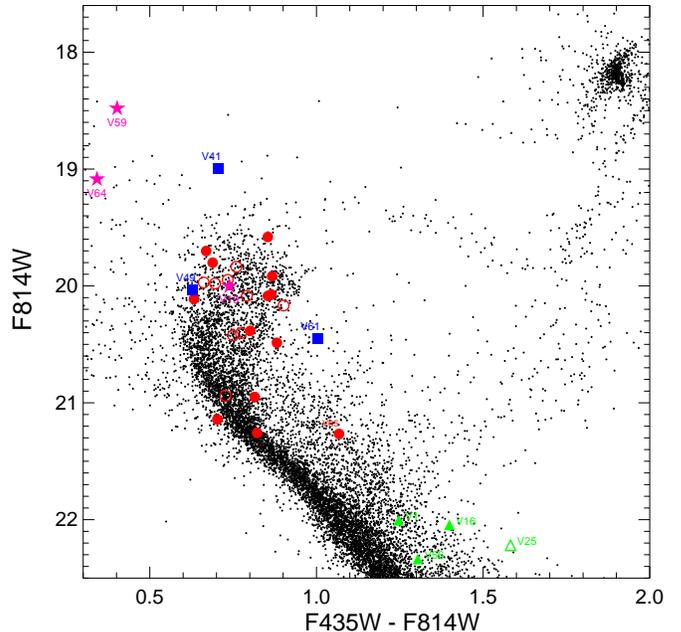}
      \caption{HST photometry of NGC 1846 from \citet{milone09} zoomed in the MSTO and upper MS area. 33 of the 36 variable stars found in this paper within the ACS fov are shown.  Open symbols represent variables with poor quality photometry as defined in \citet{milone09}, while filled symbols are variables with good measurements. Red symbols are  \dsct, blue symbols are RRL, green symbols are eclipsing binaries and cyan symbols are variables with undefined class. For clarity only a few \dsct discussed in the main text are labeled. }
         \label{fig:hst}
   \end{figure}
   
\subsection{A comparison with \textit{HST} photometry}

Even though the number of discovered \dsct, their estimated total amount, and in general the low amplitudes found are most likely not enough to produce any significant broadening of the MSTO, it is interesting to see what impact these discovered \dsct have in the published \textit{HST} photometry. 

We cross-matched the F435W/F814W \textit{HST}/ACS photometry of NGC 1846 from \citet{milone09} to the positions of our discovered variables. Given the much smaller ACS fov, only 35 of the discovered variables are found within this fov; 25 being \dsct. These can be seen as red symbols in Fig.\ref{fig:hst}. Catalogues were matched using a combination of \textsc{CataComb}$^5$\footnote{$^5$ Developed by Paolo Montegriffo at the Bologna Astronomical Observatory} and \textsc{stilts} \citep{taylor06}.

Just like in the case of the GMOS photometry (Fig.~\ref{fig:cmd}), variables stars concentrate at the MSTO. Fig.~\ref{fig:hst} shows the ACS photometry of the upper MS and MSTO areas where 32 of the discovered variables stars are. Open symbols are stars with poorly measured photometry according to \citet{milone09} based on the frame-to-frame scatter (see below), uncertainty in the frame-to-frame position and the residual of the psf fit.

One notable feature is that \dsct do not lie preferentially along any of the split sequences, but instead scatter along both and with slightly larger range in color, as expected for variable stars caught at a random phase.

The \citet{milone09} ACS photometry of NGC 1846 comes from 3 exposures in F435W and 4 in F814W. Despite that \citet{milone09} applied a selection based on the rms of the magnitudes in different exposures to set up the good quality sample, 24 out of the 35 variables were labeled as good quality by them, including all 3 RRL which are expected to have the largest rms based on their larger amplitudes. This again proves the inadequacy of using poorly time-sampled data, as is the case of all IA LMC clusters observed with \textit{HST}, in order to detect variable stars.

\section{Summary and conclusions}

In this paper we explored the influence the \dsct pulsators have in the morphology of the MSTO in IA clusters in the LMC. Since the great majority of the photometry of these clusters comes from using merely 1 or 2 images per filter \citep[e.g.][]{mackey07,milone09,piatti14}, these variables have so far being undetected, and their role ignored.

Using new time series photometry of the LMC IA cluster NGC 1846 obtained with Gemini South, we have discovered 55 \dsct in the field of NGC 1846, plus 18 variables of other types. This is the first in-depth study of the short-period variables in an LMC cluster.  Considering completeness and background contamination we estimate the number of \dsct belonging to the cluster somewhere around 40 and 60. This number of \dsct will not produce a significant impact in the MSTO morphology, as hundreds or even thousands would be needed to produce a significant broadening of the MSTO according to the modelling of \citet{salinas16b}.
 
This estimated total number of \dsct is still very uncertain. We assumed the radial distribution of \dsct follows the light of the cluster to extrapolate the density of \dsct to the inner radii where crowding makes measurements impossible. Moreover, we detected only one \dsct within the half-light radius, where a large number of variables should still be uncovered. Only time-series observations with higher spatial resolution will provide the final answer to the role \dsct have in the broadening of MSTOs.

Finally, we note that NGC 1846 is, to our knowledge,  the star cluster with the largest population of \dsct ever discovered, including open clusters in our Galaxy \citep[e.g.][]{andersen09,sandquist16}. This opens new avenues for the study of PL relations of \dsct, for example, as function of metallicity, that will be studied in a forthcoming paper. The fact that we have discovered more than 50 of these variables in a patch of the sky where OGLE detected none, is a warning that OGLE is most probably severely incomplete for fields with only mild crowding, and very biased towards variables with long periods and large amplitudes; therefore probably thousands of \dsct, fainter, mostly eclipsing, variables and even some RRL located in the distant edge of the LMC, remain to be discovered.

\acknowledgements
We thank the anonymous referee for a fast report which helped us clarify several issues. RS thanks Antonino Milone for making his \textit{HST}/ACS photometry of NGC 1846 available and Ernst Paunzen for pointing us to the Galactic clusters richest in \dsct. We thank Catherine Kaleida for all her efforts organizing the REU program at CTIO. This project was conducted in the framework of the CTIO REU Program, which is supported by the National Science Foundation under grant AST--1062976. JS acknowledges partial support from NSF grant AST--1308124 and the Packard Foundation. Based on observations made with the NASA/ESA Hubble Space Telescope, and obtained from the Hubble Legacy Archive, which is a collaboration between the Space Telescope Science Institute (STScI/NASA), the Space Telescope European Coordinating Facility (ST-ECF/ESA) and the Canadian Astronomy Data Centre (CADC/NRC/CSA).

\facility{Gemini-S}

\software{IRAF \citep{tody86,tody93}, ISIS (v 2.1, \citealt{alard00}), CataComb, stilts \citep{taylor06}, daophot/allstar/allframe \citep{stetson87,stetson94}, daomatch/daomaster \citep{stetson93}}

\appendix

\section{Notes on some individual variables}\label{sec:notes}

V1: even though this could easily be an eclipsing variable at the MSTO of the old LMC population, its curve at maximum light also resembles a RRc variable. 

V5: by its color and luminosity is most probably a MS contact binary, although its shape looks more sawtooth-like than sinusoidal.

V12: according to \citet{soszynski09a}, this is RRL OGLE-LMC-RRLYR-05278 with period 0.5870247d.
 
V19: has the magnitude and color of cluster \dsct, but its shape does not resemble a \dsct, therefore we classify it as unknown type.

V25: this is a very faint source for which we do not have psf photometry and therefore the light curve cannot be transformed into magnitudes. We originally classified this variable as \dsct based on its shape and period, but its position in the \textit{HST} CMD (Fig . \ref{fig:hst}), reveals it as a more likely eclipsing binary.

V34: another very faint source, for which we lack color information. Its light curve resembles more a \dsct than a eclipsing binary, therefore is probably a background source.

V41: according to \citet{soszynski09a}, this the RRe (second-overtone pulsator) OGLE-LMC-RRLYR-05379 with a period of 0.272447d

V48: this is OGLE-LMC-RRLYR-05394 from  \citet{soszynski09a}, with period 0.472163 d.

V49: even though it has a magnitude and color that puts it right at the MSTO, its period is much longer than the timespan of the observations. Even though it could be a \dsct with a very long period, it is more likely a background RRL that was not discovered by OGLE.

V50: another source without color. Shape of light curve and period indicate a \dsct, although its faintness implies a background source. With $A_r$=0.5 is one of the variables with the largest amplitudes in the sample.

V52: has the color of a  \dsct, but a much fainter magnitude. Probably a background \dsct.

V53: a \dsct appearing about a magnitude below the MSTO. Most likely a LMC field \dsct somewhat on the background.

V55: another variable too faint in $g$ to obtain a color. It is very faint in $i$, but has an amplitude $\sim$ 0.5 mag. Its light curve shape resembles a \dsct an a RRL, but we cannot adventure any firm classification. 

V56: Most probably a long period eclipsing binary.

V57: similar to V53.

V59: the brightest of all variables, but with a very small amplitude to be a RRL. We cannot determine a classification for this variable.

V60: too faint and red to be a \dsct. We set its class as Unknown.

V61: has the right magnitude and color to be a \dsct in NGC 1846, but its very high amplitude and rather long period points at a background RRL.

V64: is one of the bluest variables in the sample. Brighter than the MSTO, its partial light curve resembles the sinusoidal shape of a RRc, although that would be a very tentative classification.

V68: like V53 and V 57, this variable appears below the MSTO of NGC 1846. Is probably a background \dsct.

V70: the partial shape of this light curve hints at a background RRL.

V72: same as V52.

V75: another variable for which we could not transform relative flux into magnitudes. Its short period and light curve shape indicate is a \dsct.

\newpage

\begin{deluxetable*}{lcccccllc}\label{table:lcs}
\tablewidth{0pt}
\tablecaption{Positions, mean magnitudes, $r$ amplitudes, periods and classification for all the variables discovered in the NGC 1846 field. Uncertain amplitudes are indicated with a colon, while the $>$ indicates lower limits for some periods. The classification is D: delta Scuti, RRL: RR Lyrae, E: eclipsing binary, LP: long period variable, and U: variable with unknown classification. The last column indicates either a cross identification with OGLE or a note in Sect. \ref{sec:notes} indicated with the * sign.}
\tablehead{
\colhead{ID} & \colhead{RA (J2000)} & \colhead{Dec (J2000)}& \colhead{$\langle g\rangle$} & \colhead{$\langle r\rangle$} & \colhead{$A_r$} &\colhead{$P$(d)} & \colhead{Type} &  \colhead{Note}
}
\startdata
V1 & 05:07:10.061 & --67:28:34.35 & 22.60 & 22.31 & 0.31 & 0.17 & E & *\\   
V2 & 05:07:11.184 & --67:28:34.85 & 21.10 & 20.98 & 0.03 & 0.042 & D & \\   
V3 & 05:07:11.684 & --67:29:44.91 & 21.37 & 21.26 & 0.02 & 0.043 & D \\   
V4 & 05:07:13.242 & --67:27:51.07 & 20.78 & 20.66 & 0.02 & 0.076 & D \\   
V5 & 05:07:15.079 & --67:27:30.68 & 24.58 & 23.97 & 0.32 & 0.13 & E &*\\   
V6 & 05:07:15.856 & --67:30:03.18 & 21.27 & 21.05 & 0.03: & 0.14 & D \\   
V7 & 05:07:16.925 & --67:25:07.09 & 21.35 & 21.30 & 0.10 & 0.060 & D \\   
V8 & 05:07:16.946 & --67:26:52.67 & 21.06 & 20.90 & 0.04 & 0.064 & D \\   
V9 & 05:07:17.633 & --67:29:31.52 & 20.90 & 20.79 & 0.03 & 0.075 & D \\   
V10 & 05:07:19.007 & --67:28:24.38 & 21.57 & 21.42 & 0.02 & 0.052 & D \\   
V11 & 05:07:19.084 & --67:27:34.23 & 24.23 & 23.79 & 0.25: & $>0.15$ & E \\   
V12 & 05:07:20.048 & --67:29:57.74 & 19.64 & 19.38 & 0.10: & $>0.2$ & RRL &  OGLE-LMC-RRLYR-05278 \\   
V13 & 05:07:20.256 & --67:25:30.38 & 21.80 & 21.59 & 0.08 & 0.042 & D \\   
V14 & 05:07:22.364 & --67:30:26.48 & 22.09 & 21.80 & 0.03 & 0.15 & D \\   
V15 & 05:07:22.941 & --67:28:07.24 & 21.02 & 20.95 & 0.04 & 0.076 & D \\   
V16 & 05:07:23.480 & --67:27:21.83 & 22.44 & 22.01 & 0.18 & 0.13 & D \\   
V17 & 05:07:23.651 & --67:30:04.94 & 21.69 & 21.48 & 0.03 & 0.054 & D \\   
V18 & 05:07:24.485 & --67:27:08.23 & 20.98 & 20.69 & 0.03 & 0.14 & D \\   
V19 & 05:07:24.787 & --67:27:27.30 & 20.51 & 20.42 & 0.04: & $>0.14$ & U & * \\   
V20 & 05:07:25.334 & --67:26:53.16 & 20.12 & 20.00 & 0.02 & 0.1 & D \\   
V21 & 05:07:25.431 & --67:26:37.37 & 21.59 & 21.46 & 0.05 & 0.10 & D \\   
V22 & 05:07:27.099 & --67:25:24.06 & 20.36 & 20.33 & 0.03: & 0.12 & D \\   
V23 & 05:07:27.237 & --67:27:02.63 & 20.35 & 20.28 & 0.04 & 0.068 & D \\   
V24 & 05:07:28.438 & --67:26:11.55 & 20.69 & 20.51 & 0.03: & 0.10 & D \\   
V25 & 05:07:30.025 & --67:28:18.75 & ---      & --- & --- & 0.13 & E & * \\   
V26 & 05:07:30.772 & --67:25:42.96 & 22.07 & 21.76 & 0.06 & 0.046 & D \\   
V27 & 05:07:31.283 & --67:26:12.13 & 20.49 & 20.46 & 0.01 & 0.044 & D \\   
V28 & 05:07:31.454 & --67:29:44.21 & 22.59 & 22.41 & 0.09 & 0.046 & D \\   
V29 & 05:07:31.953 & --67:25:14.02 & 22.94 & 22.49 & 0.12: & $>0.12$ & E \\   
V30 & 05:07:32.221 & --67:25:11.77 & 22.22 & 21.93 & 0.08 & 0.15 & D \\   
V31 & 05:07:32.974 & --67:28:19.35 & 20.51 & 20.42 & 0.02 & 0.074 & D \\   
V32 & 05:07:33.023 & --67:26:56.07 & 20.55 & 20.43 & 0.04 & 0.078 & D \\   
V33 & 05:07:33.096 & --67:30:01.91 & 20.57 & 20.45 & 0.20: & 0.11 & D \\   
V34 & 05:07:33.336 & --67:25:32.26 & ---      & 24.09 & 0.25 & 0.12 & U & * \\   
V35 & 05:07:33.519 & --67:28:22.44 & 20.57 & 20.44 & 0.08 & 0.12 & D \\   
V36 & 05:07:34.001 & --67:28:16.31 & 20.24 & 20.21 & 0.02 & 0.078 & D \\   
V37 & 05:07:34.180 & --67:25:48.84 & 21.74 & 21.77 & 0.13 & 0.14 & D \\   
V38 & 05:07:35.155 & --67:28:38.01 & 20.70 & 20.59 & 0.03 & 0.125 & D \\   
V39 & 05:07:35.255 & --67:26:20.48 & 20.39 & 20.33 & 0.02 & 0.071 & D \\   
V40 & 05:07:36.298 & --67:26:11.30 & 22.33 & 22.14 & 0.10 & 0.14 & E \\   
V41 & 05:07:36.410 & --67:30:20.05 & 19.49 & 19.42 & 0.07: &$ >0.16$ & RRL & OGLE-LMC-RRLYR-05379\\   
V42 & 05:07:36.430 & --67:25:32.14 & 20.73 & 20.70 & 0.02 & 0.038 & D \\   
V43 & 05:07:37.626 & --67:28:21.32 & 20.63 & 20.50 & 0.02 & 0.095 & D \\   
V44 & 05:07:37.799 & --67:28:33.29 & 20.47 & 20.39 & 0.10 & 0.075 & D \\   
V45 & 05:07:37.816 & --67:27:12.63 & 20.63 & 20.46 & 0.03 & 0.104 & D \\   
V46 & 05:07:38.220 & --67:26:53.09 & 20.74 & 20.74 & 0.04 & 0.071 & D \\   
V47 & 05:07:38.335 & --67:25:36.50 & 21.29 & 21.24 & 0.08: & $>0.20$ & D \\   
V48 & 05:07:38.412 & --67:25:56.26 & 20.06 & 19.80 & 0.08: & $>0.16$ & RRL &OGLE-LMC-RRLYR-05394 \\   
V49 & 05:07:38.618 & --67:27:53.52 & 20.55 & 20.50 & 0.16: & $>0.2$ & RRL& * \\   
V50 & 05:07:38.767 & --67:25:09.38 & ---      & 24.44 & 0.50 & 0.10 & U& * \\   
V51 & 05:07:39.109 & --67:29:28.91 & 20.20 & 20.02 & 0.02 & 0.15 & D \\   
V52 & 05:07:39.754 & --67:28:36.16 & 23.72 & 23.76 & 0.40 & 0.071 & D & * \\   
V53 & 05:07:39.796 & --67:25:43.87 & 22.13 & 22.17 & 0.15 & 0.16 & D & *\\   
V54 & 05:07:42.020 & --67:30:02.59 & 21.68 & 21.56 & 0.02 & 0.049 & D \\   
V55 & 05:07:42.945 & --67:26:05.89 & ---      & 23.42 & 0.50: & $>0.1$ & U& * \\   
V56 & 05:07:43.251 & --67:29:03.09 & 23.19 & 22.87 & 0.25: & $>0.15$ & E & *\\   
V57 & 05:07:45.914 & --67:28:30.28 & 21.80 & 21.86 & 0.05 & 0.049 & D & * \\   
V58 & 05:07:45.974 & --67:25:53.99 & 20.51 & 20.47 & 0.03 & 0.11 & D \\   
V59 & 05:07:46.526 & --67:27:41.14 & 18.78 & 18.86 & 0.01 & $>0.16$ & U & *\\   
V60 & 05:07:46.679 & --67:25:40.43 & 22.55 & 22.26 & 0.11 & $>0.10$ & U & * \\   
V61 & 05:07:47.170 & --67:28:22.70 & 21.31 & 21.15 & 0.80: & $>0.2$ & RRL & * \\   
V62 & 05:07:48.396 & --67:25:39.26 & 20.05 & 19.95 & 0.11 & 0.09 & D \\   
V63 & 05:07:49.025 & --67:25:54.81 & 20.77 & 20.87 & 0.12 & 0.07 & D \\   
V64 & 05:07:49.655 & --67:28:25.60 & 19.34 & 19.45 & 0.01: & $>0.16$ & U & *\\   
V65 & 05:07:50.038 & --67:28:11.99 & 22.05 & 21.79 & 0.06 & $>0.16$ & D \\   
V66 & 05:07:50.236 & --67:28:22.26 & 20.57 & 20.55 & 0.01 & 0.047 & D \\   
V67 & 05:07:53.337 & --67:28:42.72 & 23.53 & 23.04 & 0.35 & 0.11 & E \\   
V68 & 05:07:53.811 & --67:27:21.08 & 21.72 & 21.80 & 0.12: &$ >0.16$ & D & *\\   
V69 & 05:07:54.035 & --67:27:55.26 & 20.65 & 20.54 & 0.02 & 0.127 & D \\   
V70 & 05:07:54.560 & --67:29:40.68 & 21.98 & 21.77 & 0.12: & $>0.2$ & RRL &  *\\   
V71 & 05:07:54.581 & --67:27:57.52 & 21.16 & 21.08 & 0.02 & 0.067 & D \\   
V72 & 05:07:55.982 & --67:28:30.90 & 23.61 & 23.57 & 0.34 & 0.128 & D & *\\   
V73 & 05:07:56.233 & --67:26:10.71 & 22.16 & 22.03 & 0.05 & 0.15 & D \\   
V74 & 05:07:56.857 & --67:27:46.59 & 21.06 & 20.96 & 0.02 & 0.057 & D \\   
V75 & 05:07:59.875 & --67:25:28.64 & --- & --- & --- & 0.095 & D & *\\   
V76 & 05:08:00.985 & --67:28:55.19 & 22.88 & 22.53 & 0.12: & $>0.18$ & E  
\enddata
\label{table:variables}
\end{deluxetable*}

   \begin{figure*}
   \centering
   \includegraphics[width=\textwidth]{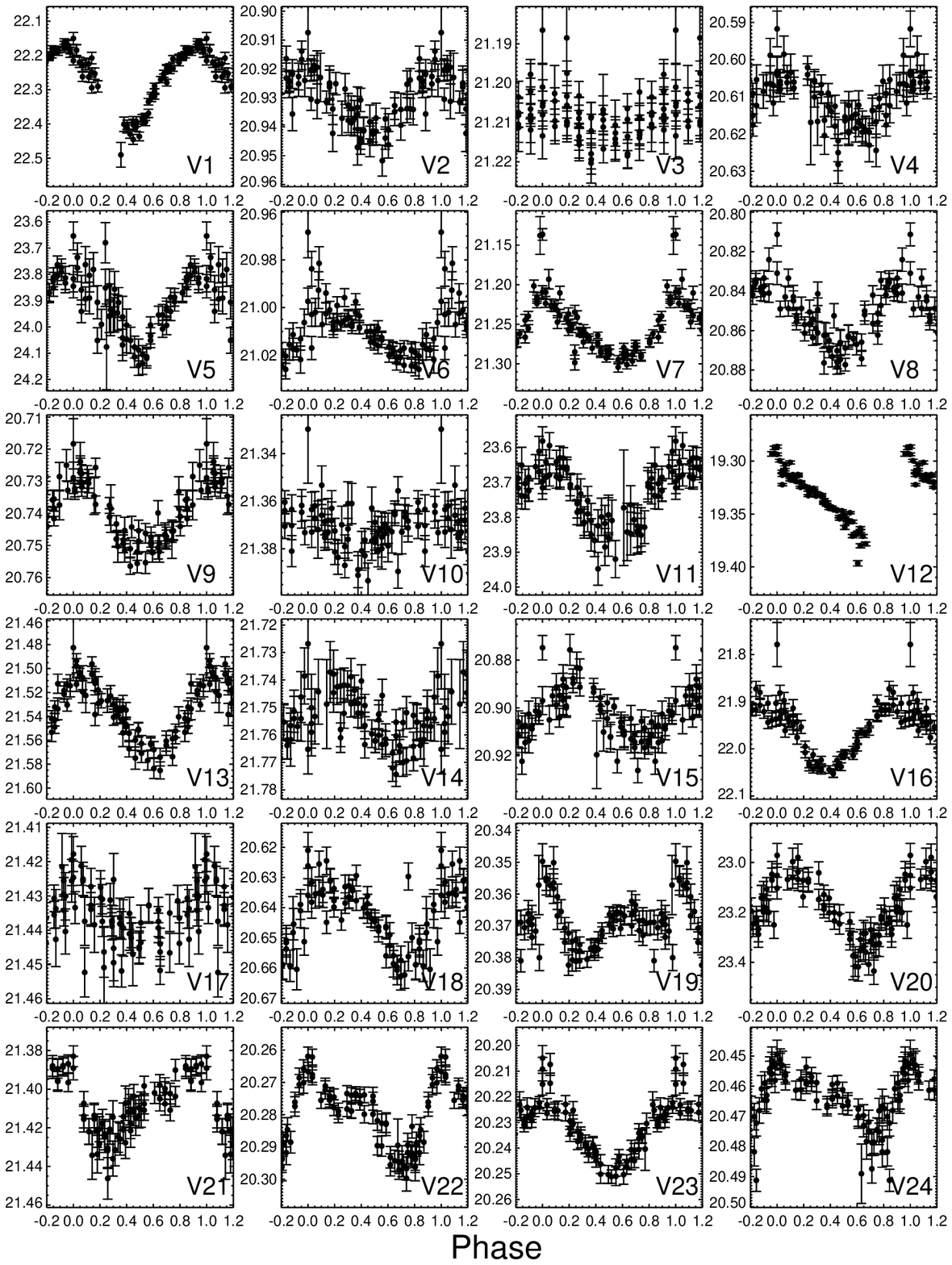}
      \caption{Folded light curves of all the new variables stars in the NGC 1846 field. All light curves were measured in the $r$ band. Note that a couple of them (V25 and V75) were not calibrated into standard magnitudes and their light curve are only shown in relative counts.  Variables where only a minimum period is given in Table \ref{table:lcs}, were phased using this minimum period.}
               \label{fig:variables1}
   \end{figure*}

   \begin{figure*}
   \centering
   \includegraphics[width=\textwidth]{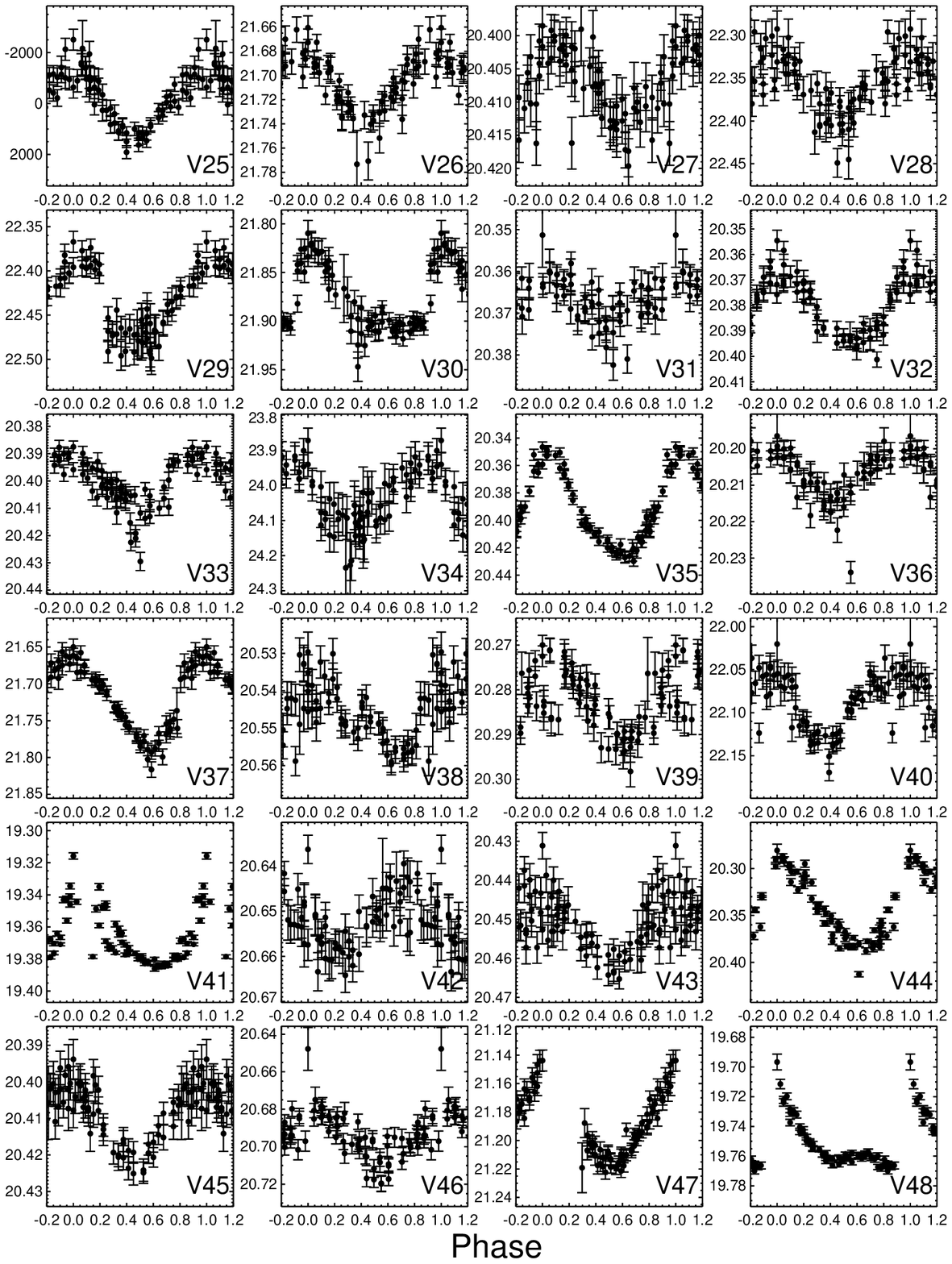}
      \caption{Same as in Fig. \ref{fig:variables1}.}
               \label{fig:variables2}
   \end{figure*}
   
     \begin{figure*}
   \centering
   \includegraphics[width=\textwidth]{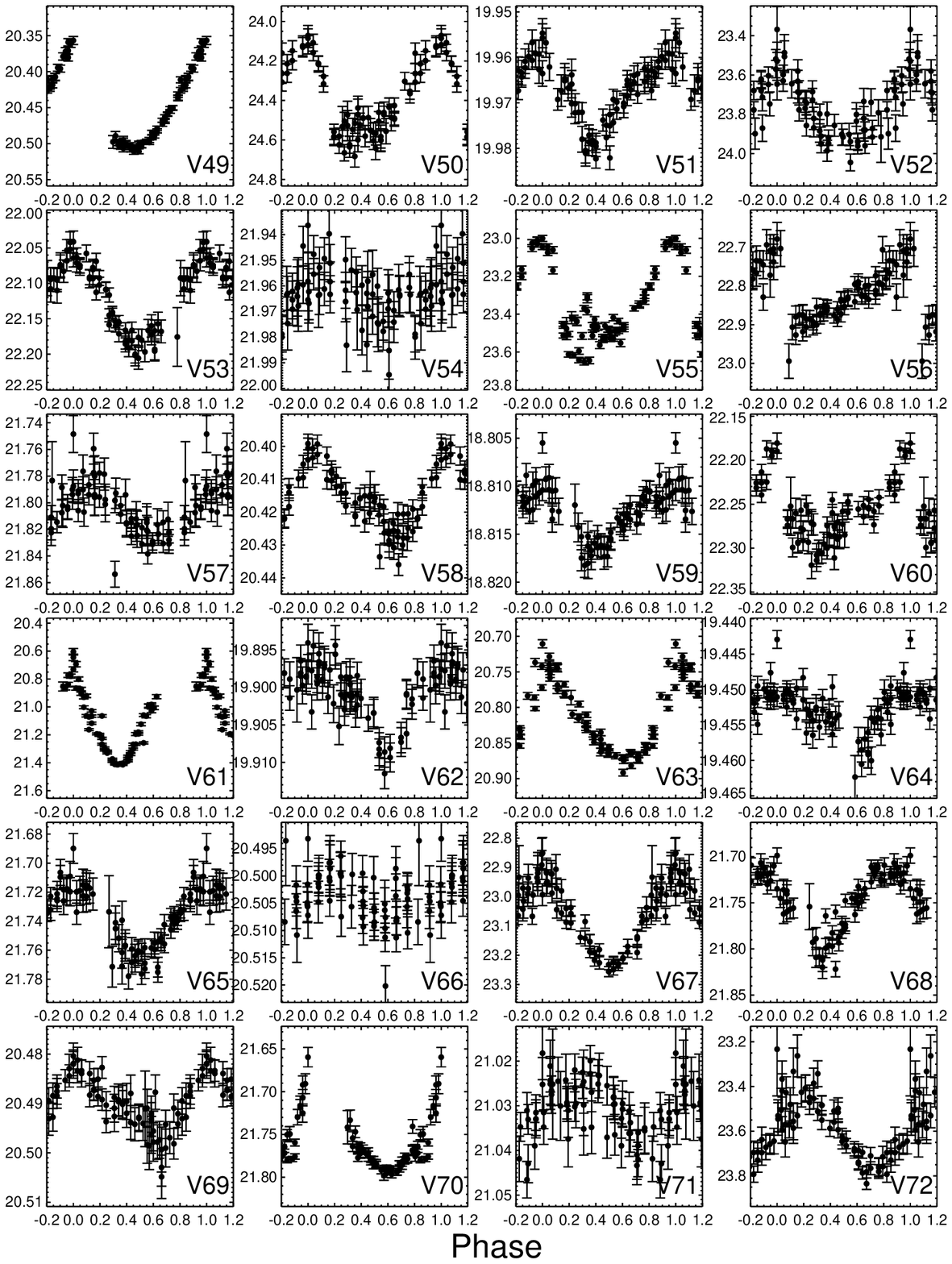}
      \caption{Same as in Fig. \ref{fig:variables1}.}
               \label{fig:variables3}
   \end{figure*}
   
       \begin{figure*}
   \centering
   \includegraphics[width=\textwidth]{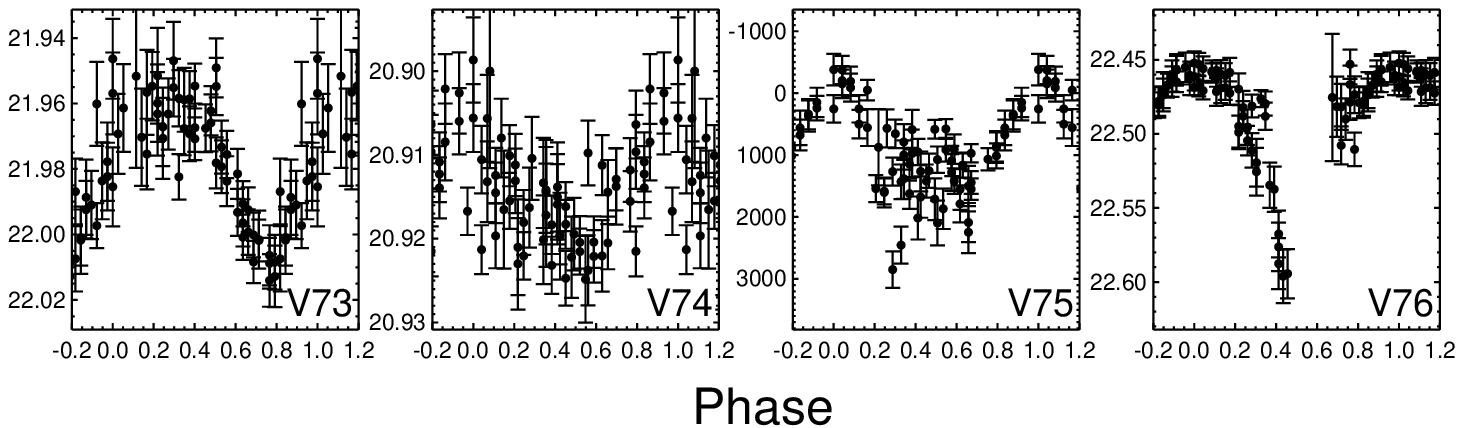}
      \caption{Same as in Fig. \ref{fig:variables1}.}
               \label{fig:variables4}
   \end{figure*}
   
          \begin{figure*}
   \centering
   \includegraphics[width=0.9\textwidth]{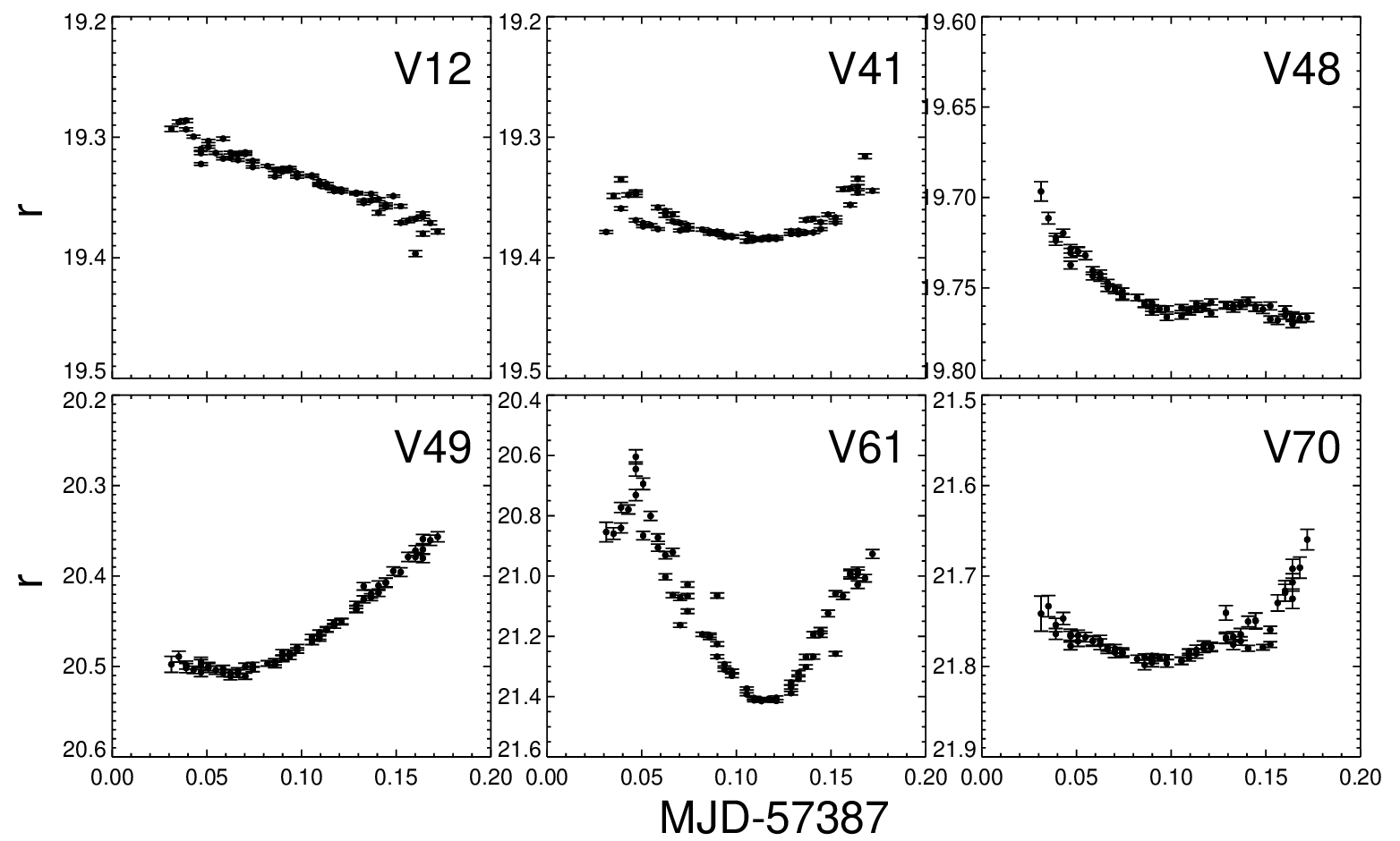}
      \caption{Known RRL (top row) discovered by OGLE and new candidate RRL (bottom row). Light curves are given in modified Julian date vesus magnitude.}
               \label{fig:rrl}
   \end{figure*}
   
\end{document}